\documentclass[aps,pra,showpacs,twocolumn,superscriptaddress]{revtex4}
\usepackage{graphicx,bbm,amsthm, amsmath, amssymb}
\usepackage{color}
\def\a{\gamma}
\newcommand{\ket}[1]{| #1\rangle}

\begin{document}
\title{Multifractality of quantum wave packets}
\author{Ignacio Garc\'{\i}a-Mata}
\affiliation{Instituto de Investigaciones F\'isicas de Mar del Plata
(IFIMAR), CONICET--UNMdP,
Funes 3350, B7602AYL
Mar del Plata, Argentina.}
\affiliation{Consejo Nacional de Investigaciones Cient\'ificas y
Tecnol\'ogicas (CONICET), Argentina}
\author{John Martin}
\affiliation{Institut de Physique Nucl\'eaire, Atomique et de
Spectroscopie, Universit\'e de Li\`ege, B\^at.\ B15, B - 4000
Li\`ege, Belgium}
\author{Olivier Giraud}
\affiliation{LPTMS, CNRS and Universit\'e Paris-Sud, UMR 8626, B\^at. 100,
91405 Orsay, France}
\author{Bertrand Georgeot}
\affiliation{%
Universit\'e de Toulouse; UPS; Laboratoire de
 Physique Th\'eorique (IRSAMC); F-31062 Toulouse, France
}
\affiliation{CNRS; LPT (IRSAMC); F-31062 Toulouse, France}
\date{\today}%
\begin{abstract}
We study a version of the mathematical Ruijsenaars-Schneider model, and reinterpret it physically in order to describe the spreading with time of quantum wave packets in a system where multifractality can be tuned by varying a parameter. We compare different methods to measure the multifractality of wave packets, and identify the best one.  We find the multifractality to decrease with time until it reaches an asymptotic limit, different from the multifractality of eigenvectors, but related to it, as is the rate of the decrease. Our results could guide the study of experimental situations where multifractality is present in quantum systems.
\end{abstract}
\maketitle

\section{Introduction}
Multifractal properties  have been characterized in several physical
contexts, from turbulence  \cite{turbulence} to the stock market \cite{stock}
or cloud images \cite{cloud}.
Similar features were also recently observed in quantum mechanics 
or complex wave systems. Indeed,  
multifractal wave functions are observed for
electrons at the Anderson
metal-insulator transition \cite{trans,mirlin2000,mirlinRMP08,romer}, 
in quantum Hall transitions \cite{huckenstein}, in Random Matrix models \cite{PRBM,ossipov} and others \cite{indians1,indians,garciagarcia}.
They are also visible 
in the different context of pseudointegrable systems, for which constants
of motion exist, but dynamics takes place
on manifolds more complicated than the tori characteristic of integrability \cite{interm,giraud,bogomolny,MGG,MGGG,BogGirPRL,BogGir11,BogGir12}.

Many theoretical studies have been devoted  to these quantum multifractal systems. 
In parallel, experimental progress opens the way to direct observation of multifractality; hints of such properties were seen 
in waves in elastic media \cite{billes}, disordered conductors \cite{richard} and cold atoms \cite{cold}. The theoretical studies were mainly
concentrated on eigenvectors (stationary states) of the systems considered \cite{mirlin2000,mirlinRMP08,romer,huckenstein,PRBM,ossipov,indians1,indians,garciagarcia,interm,giraud,bogomolny,MGG,MGGG,BogGirPRL,BogGir11,BogGir12}. In contrast, the experimental protocols in general involve the propagation of Wave Packets (WP), for which results on eigenvectors are a priori not directly applicable.
In order to further characterize
experimental results and interpret them, it is therefore important 
to have detailed results on the multifractal properties of WP.
Some works have related global properties
of WP, e.g.~spreading laws or envelope shapes, to
the multifractal properties of eigenvectors or eigenspectra \cite{chalker,schweitzer,guarneri,geisel,raizen}. Other works found specific examples of multifractal
WP \cite{schreiber,grecs}. However, the general existence and origin of the multifractality of WP  is still unclear.  In this paper, we reinterpret the mathematical Ruijsenaars-Schneider model \cite{ruijsc}
 as the quantization of a pseudo-integrable map. The properties of the eigenfunctions can be continuously tuned through system parameters from a weak to a strong multifractality regime, enabling us
to systematically compare multifractality of WP and eigenvectors.  Although the system is of mathematical origin, it can serve as a testbed and the results for this model can give insights for the behavior of a wide class of physical systems with multifractal properties.
Our computations show that several numerical methods can give different results for measuring this multifractality of WP, and we identify the optimal one. Our numerical and analytical results show that one can
systematically relate the multifractality of WP and its time evolution
to the one
of eigenfunctions, opening the possibility to probe these properties in detail
through experimental observations.  

\section{The model}
 
We consider a periodically kicked system with period $T$ and Hamiltonian $H(p,q)=\frac{p^2}{2}+V(q)\sum_n\delta(t-nT)$,
with potential $V(q)=-\a \{q\}$. Here $\{q\}$ denotes the fractional part of $q$, and $(p,q)$ are the conjugated momentum and position variables.  The classical equations of motion integrated over one period yield the classical map
$ \bar{p} = p + \a ;  \;\;\;
\bar{q} = q+ T\bar{p}  \;\;(\mathrm{mod} \;1).$ The quantization 
of this map gives the unitary evolution operator 
$\hat{U}=\mathrm{e}^{- iT  
\hat{p}^2/(2\hbar)}\mathrm{e}^{- iV(\hat{q})/\hbar}\ .$  In  
\cite{giraud, MGGG}, the choice of parameters led to a quantum map on a  
toroidal phase space independent on the Hilbert space dimension $N$. 
In order to allow for long  
spreading times for a WP, here we fix $\hbar$ and truncate  
the phase space by taking $p\in[0,2\pi N\hbar[$, or equivalently  
integer indices $P$ defined by $p=2\pi P\hbar$ such that $0\leq P\leq  
N-1$. This defines a quantum map over a phase space whose classical  
size grows with $N$. The evolution operator then becomes
$\hat{U}=\mathrm{e}^{- i\pi T  
P^2}F^{-1}\mathrm{e}^{2\pi i\a Q/N}F\ ,$ with $F_{PQ}=\exp(i P  
Q/N)/\sqrt{N}$, which yields
\begin{equation}
\label{ruij}
    U_{PP'}=\frac{e^{i\Phi_P}}{N}\frac{1-e^{2i\pi \a}}{1-e^{2i\pi (P-P'+\a)/N}}\ ,
\end{equation}
with $\Phi_P=-\pi T P^2$.

This system corresponds to the mathematical Ruijsenaars-Schneider map \cite{BogGir11,BogGir12}.  This
model reinterpreted physically in this way has many advantages. The multifractality of eigenvectors is known \cite{BogGirPRL,BogGir11,BogGir12} to depend on the parameter $\gamma$ in a continuous way, enabling
to probe all the regimes from weak to strong multifractality.  In addition, the
simplicity of this 1D model makes analytical calculations and numerical computations
tractable. In the numerical results below, we replaced the kinetic
term $\Phi_{P}$ in (\ref{ruij}) by random phases, in order to get averaged quantities
while keeping the same physics.

\section{Analytical calculation of average WP}
We consider the evolution of a WP initially localized on one single momentum state, $\Psi_P^{(0)}=\delta(P-P_0)$. Iterations of the map make the WP spread out.  Analytical calculations are possible in the regime of small $t$ and $\gamma$ close to an integer. In this section, using a tailored version of perturbation theory, we compute the average WP over random phases $\Phi_P$.

A perturbation expansion for the matrix \eqref{ruij} can be obtained whenever $\a$ is close to an integer, namely $\a=k+\epsilon$ with $k$ an integer. In order to obtain slightly simpler expressions we rescale the matrix $U_{PP'}$ by a trivial factor $\exp(-i\pi\epsilon(1-1/N))$, so that it can be expressed as 
\begin{eqnarray}
 U_{PP'}&=&\mathrm{e}^{i\Phi_P}\delta(P+k-P')\nonumber\\
&&-\epsilon\frac{2\pi i}{N}\mathrm{e}^{i\Phi_P}\frac{1-\delta(P+k-P')}{1-\mathrm{e}^{2\pi i(P+k-P')/N } }+\mathcal{O}(\epsilon^2).
\label{matrix_k}
\end{eqnarray}
We consider the evolution of a wave packet initially in the state $\ket{\Psi^{(0)}}$. Upon one iteration of the map \eqref{matrix_k} the state becomes $\ket{\Psi^{(1)}}$ with components
\begin{equation}
\Psi^{(1)}_P =\mathrm{e}^{i\Phi_P}\Psi^{(0)}_{P+k}-\epsilon\frac{2\pi i}{N}\sum_{P'\neq P}\frac{\mathrm{e}^{i\Phi_P}}{1-\mathrm{e}^{2\pi i(P-P')/N }}\Psi^{(0)}_{P'+k}
\label{state1}
\end{equation}
(all indices are to be understood modulo $N$). The iterate $\ket{\Psi^{(t)}}$ after $t$ applications of the map is obtained by applying \eqref{state1} recursively. The general term at first order is of the form
\begin{equation}
\Psi^{(t)}_P =\mathrm{e}^{i\Phi_P^{(t)}}\Psi^{(0)}_{P+kt}-\epsilon\frac{2\pi i}{N}\sum_{P'\neq P}\frac{\chi_{P,P'}^{(t)}}{1-\mathrm{e}^{2\pi i(P-P')/N }}\Psi^{(0)}_{P'+kt},
\label{statet}
\end{equation}
In particular for $t=1$, $\Phi_P^{(1)}=\Phi_P$ and $\chi_{P,P'}^{(1)}=\mathrm{e}^{i\Phi_P}$.
Applying one iteration to state \eqref{statet} yields 
\begin{eqnarray}
&\Psi^{(t+1)}_P =\mathrm{e}^{i\Phi_P+i\Phi_{P+k}^{(t)}}\Psi^{(0)}_{P+k(t+1)} \nonumber \\
 &-\epsilon\frac{2\pi i}{N}\sum_{P'\neq P}\frac{\mathrm{e}^{i\Phi_P}\chi_{P+k,P'+k}^{(t)}+\mathrm{e}^{i\Phi_P+i\Phi_{P'+k}^{(t)}}}{1-\mathrm{e}^{2\pi i(P-P')/N }}\Psi^{(0)}_{P'+k(t+1)},
\end{eqnarray}
so that $\Phi_P^{(t)}$ and $\chi_{P,P'}^{(t)}$ satisfy the recurrence relations 
\begin{equation}
\Phi_P^{(t+1)}=\Phi_{P+k}^{(t)}+\Phi_P
\end{equation}
and
\begin{equation}
\chi_{P,P'}^{(t+1)} = \mathrm{e}^{i\Phi_P}\chi_{P+k,P'+k}^{(t)}+\mathrm{e}^{i\Phi_P+i\Phi_{P'+k}^{(t)}}\ .
\end{equation}
We readily obtain the following expressions
\begin{equation}
\Phi_P^{(t)}=\sum_{j=0}^{t-1}\Phi_{P+k j}
\end{equation}
and 
\begin{equation}
\chi_{P,P'}^{(t)}=\sum_{r=1}^{t}\exp\left(i\sum_{j=0}^{r-1}\Phi_{P+k j}+i\sum_{j=r}^{t-1}\Phi_{P'+k j}\right),
\label{chimn}
\end{equation}
which together with \eqref{statet} give the first-order expression of $\ket{\Psi^{(t)}}$. Now suppose we start from a wavepacket initially localized at $P_0$, so that the initial state is defined by $\Psi^{(0)}_{P_0}=1$ and its other components equal to zero. In \eqref{statet}, only terms with $P'+k t=P_0$ yield a nonzero contribution, so that we get at lowest order
\begin{eqnarray}
|\Psi_P^{(t)}|^2&=&1\ \ \ \ \ \ \ \textrm{if } P=P_0-kt\\
|\Psi_P^{(t)}|^2&=&\frac{\epsilon^2\pi^2}{N^2}\frac{|\chi_{P,P_0-kt}^{(t)}|^2}{\sin^2\frac{\pi}{N}(P-P_0+k t)} \ \ \textrm{otherwise.}
\end{eqnarray}
The first-order expression for the mean wave packet is obtained by averaging over random phases $\Phi_P$. Using \eqref{chimn}, the average reads
\begin{eqnarray}
\label{chimoy}
\langle|\chi_{P,P_0-kt}^{(t)}|^2\rangle &=&\sum_{r,r'=1}^{t}\langle\exp\Bigg[i\bigg(\sum_{j=0}^{r-1}\Phi_{P+k j}+\sum_{j=r-t}^{-1}\Phi_{P_0+k j} \nonumber \\
& &-\sum_{j=0}^{r'-1}\Phi_{P+k j}-\sum_{j=r'-t}^{-1}\Phi_{P_0+k j}\bigg)\Bigg]\rangle .
\end{eqnarray}
For diagonal terms with $r=r'$ the term in the exponential vanishes. For terms such that $r>r'$, the term in the exponential is
\begin{equation}
\label{termexp}
\sum_{j=r'}^{r-1}\Phi_{P+k j}-\sum_{j=r'-t}^{r-t-1}\Phi_{P_0+k j}.
\end{equation}
The average over random phases is nonzero if and only if all terms in \eqref{termexp} vanish, that is, if the set of indices $\Omega_1=\{P+k r',\ldots,P+k(r-1)\}$ is equal (modulo $N$) to the set $\Omega_2=\{P_0+k (r'-t),\ldots,P_0+k(r-t-1)\}$. In the case $k=0$ this is impossible since we are in the case where $P\neq P_0-kt$. Consider now $k\geq 1$ (for simplicity we restrict ourselves to $k$ coprime with $N$). Suppose that $P+k r'$ is equal to a certain index of $\Omega_2$, say $P+k r'=P_0+k (r'-t+q)$ for some $q$ with $1\leq q\leq r-r'-1$ (given that we are in a case where $P\neq P_0-kt$ we must have $q\neq 0$). Then for $0\leq s\leq r-r'-1-q$ we have equalities $P+k(r'+s)=P_0+k (r'-t+q+s)$, the last equality being $P+k(r-1-q)=P_0+k(r-t-1)$. Then in order to have $\Omega_1=\Omega_2$ we must have $P+k(r-q)$ equal to one of the remaining indices of $\Omega_2$, that is, $P+k(r-q)=P_0+k (r'-t+s_0)$ for some $s_0$, $0\leq s_0\leq q-1$. Since by definition of $q$ we have $P=P_0+k(-t+q)$, we get $k(r-r'-s_0)=0$ modulo $N$. Since we assumed that $k$ and $N$ are coprime and $r>r'$, this gives $s_0=r-r'$. But we had $s_0\leq q-1\leq r-r'-2$, which yields a contradiction. So we cannot have $\Omega_1=\Omega_2$. Thus only diagonal terms survive in \eqref{chimoy}. Since there are $t$ of them we get the final formula 
 for the average WP for $P\neq P_0-k t$:
\begin{equation}
\langle|\Psi_P^{(t)}|^2\rangle =\frac{\epsilon^2\pi^2t}{N^2}\frac{1}{\sin^2\frac{\pi}{N}(P-P_0+k t)}.
\label{bigres}
\end{equation}

It implies that close to integer values of $\gamma$, the WP
displays a single peak moving at speed $k$.  Actually, formula (\ref{bigres})
is close to the numerical results  even for quite large values of
 $t$ and for $\gamma$ far from integers, provided $k$ is replaced by $\gamma$ and $\epsilon$ by $\sin (\pi\gamma)/\pi$. This can be seen for instance on the insets of Fig.~\ref{fig1}, where the average wave function is shown together with the formula (\ref{bigres}) for three different values of $\gamma$ and $t=100$.  Discrepancies can be visible only by zooming close to the center of the distribution.  In the case $\gamma \approx 0.5$, one can actually distinguish two peaks,
one staying at the initial position, the other one moving faster than $\gamma$,
but the tails are still perfectly reproduced by Eq.(\ref{bigres}). The peak at the origin for small $\gamma$ can be interpreted as a manifestation of strong multifractality, since it can be related to the correlation dimension \cite{chalker,schweitzer,geisel}. 

Several regimes can be characterized in the evolution of the WP.
At $t=0$, the WP is localized at $P=P_0$. As time increases, the WP spreads according to Eq.(\ref{bigres}) until it reaches the system size.  We have checked that the limiting result for very long times ($t=\infty$ limit) is equivalent to the one obtained by diagonalizing the evolution operator, replacing the eigenphases by random numbers, and transforming back to the momentum basis. The speed at which the properties converge to this asymptotic regime depends on the multifractality (see section \ref{results}).

\section{Numerical computation of multifractal exponents}
Different methods can be used in order to compute multifractal exponents. They are all equivalent for mathematically defined multifractal measures. However, our system is discrete and cannot reach arbitrarily small scales.  Besides, WP can be described as a smooth average Eq.(\ref{bigres}), with superimposed fluctuations.  Our study focuses on the multifractality of these fluctuations. However, the presence of a nontrivial envelope can affect the computation of the exponents. Thus in our case different methods may give different answers.  We tested four different algorithms. The first one (moment method), widely used in a quantum context, e.g. in \cite{mirlin2000,mirlinRMP08}, consists in computing the moments of the wave function $\mathcal{P}_q= \sum_{P=1}^N |\Psi_P|^{2q}$
for different system sizes $N$; the multifractal exponent $D_q$ can be obtained from the scaling of these moments with $N$ through  $\langle \mathcal{P}_q \rangle \propto N^{-\tau_q}$ with $\tau_q=D_q(q-1)$.  However, this method assumes scale invariance of the system as $N$ increases. In our system the envelope has a nontrivial scaling
with $N$, and this effect is hard to disentangle from the multifractality
due to the fluctuations. Thus while in most regimes we found this method to give results equivalent to other ones, in some cases it gives nontrivial multifractal exponents even for a smooth WP (e.g. for the average WP of Eq.(\ref{bigres})). 

\begin{figure}[h!]
\includegraphics[width=0.95\linewidth,angle=0]{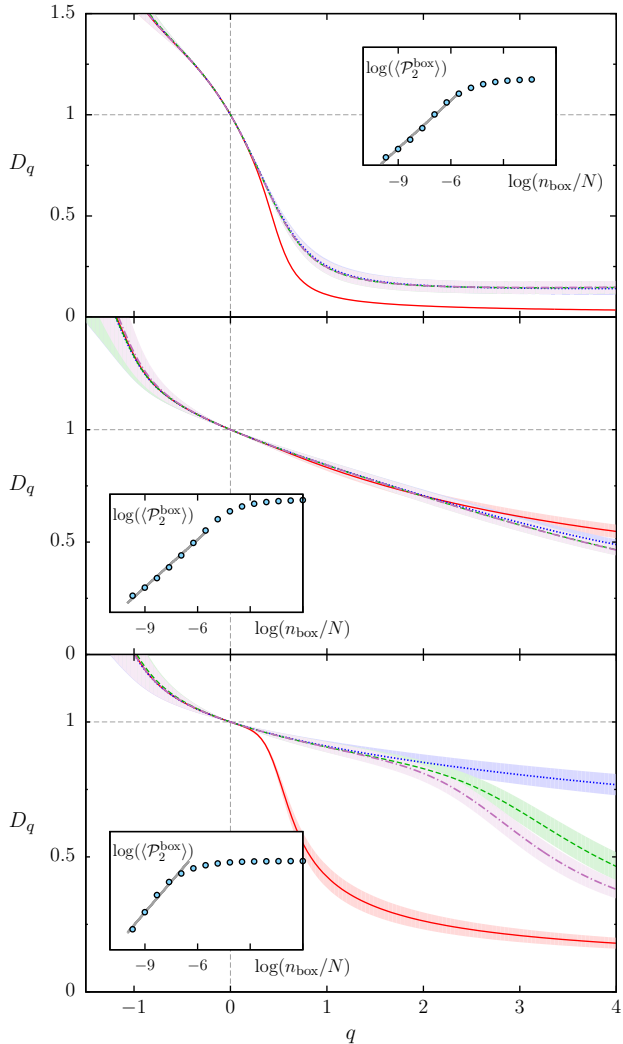}
\caption{(Color online) Multifractal exponents $D_q$ vs $q$ for $\gamma=0.05$
(top), $\gamma=0.5$ (middle) and $\gamma=0.95$ (bottom).
Red solid curves: WP at $t=100$ (including envelope and fluctuation effects, see text); blue dotted curves: fluctuations 
corresponding to the WP divided by the average WP over all realizations;
green dashed curves: fluctuations corresponding to the WP divided by a
smoothed average WP; purple dotted-dashed curves: fluctuations
corresponding to the WP divided by the analytical average WP Eq.(\ref{bigres})
(see text). In the top and middle panels the three latter curves are mostly
indistinguishable. Shaded areas indicate standard error in the
least-square fit.
Insets: examples of fit for the WP and $q=2$. The $D_q$ have been
extracted from $1000$ random phase realizations of size  $N=2^{16}$ by the
BC method applied on the $2^{15}$ central components with box sizes
ranging from $1$ to $64$. All quantities on the figure are dimensionless.
\label{fig0}}
\end{figure}

We therefore investigated alternative methods, 
which use only one system size $N$. 
One possibility is to evaluate the scaling with box size of the moments $\mathcal{P}_q^{\mathrm{box}} =\sum_{\mathrm{boxes}} (\sum_{P \in \mathrm{box}} |\Psi_P|^{2})^{q}$ of the wave function summed up inside boxes of same sizes (box-counting (BC) method) \cite{MGG,MGGG}. Another method uses  the scaling of the sum of the local maxima of the wavelet transform of the wave function at each scale (wavelet method) \cite{arneodo,kantel,nous}. Finally, we investigated a method suited for WP spreading with time, similar to the moment method but using the scaling of the moments as a function of time instead of different system sizes (time method) \cite{grecs}. We found this latter method difficult to implement since it required some knowledge of the spreading law, and less reliable than the BC method. For relatively large values of $q$
  ($q>1$) results were similar to those obtained by the other methods.
For strong multifractality, BC and
wavelet methods gave the same results.  However, in the weak multifractality
regime, where the scaling is strongly dependent on the scales at which
we choose to fit the data, the BC method appeared to be more reliable.  We therefore choose throughout the paper the BC method for the numerical computation of the exponents. An average was made over the positions of the box centers, to eliminate a threshold effect linked to the relative position of the WP and the boxes.   
In addition, it is interesting to note that the multifractality of WP contains effects of the envelope of the WP and effects from the fluctuations around this envelope. In Fig.~\ref{fig0} we show the multifractal exponents $D_q$ computed by the BC method in four different ways.  The first one is the direct application of the BC method for the full WP, the other three aim at separating fluctuation effects from envelope effects by dividing the WP by its average value, computed in three different ways (see caption of Fig.~\ref{fig0}). The results show that if $\gamma$ is not close to one, the multifractality measured by the BC method on the full WP corresponds mainly to fluctuation effects. In contrast, for $ \gamma$ approaching one (weak multifractality regime), the results of the BC method clearly incorporate both envelope and fluctuation effects.  In many physical systems, averages over realizations or analytical envelopes might be difficult to obtain, so we will use in the following computations the BC method for the full WP without dividing the WP by its average value.  

We recall that the exponents $D_q$ are positive and decrease for $q>0$ from $D_0=1$; at a fixed value of $q>0$ the smaller $D_q$ is the stronger the
multifractality is. For an ergodic wave function one has $D_q=1$ for all $q$.
In systems such as ours where an average is made over wave functions one can distinguish two
sets of multifractal exponents \cite{mirlin2000,mirlinRMP08}.  For the BC method, the first one corresponds to $\langle  \mathcal{P}_q^{\mathrm{box}} \rangle$ as defined above, and yields exponents $D_q$, while the other one corresponds to $\langle \ln  \mathcal{P}_q^{\mathrm{box}} \rangle$, and yields exponents $D_q^{\mathrm{typ}}$.  In cases where moments are distributed according to power laws with small exponents, 
the two quantities can be different, the latter being the typical value of the moment 
for the bulk of the wave functions considered, while the former 
could be dominated by rare wave functions with much larger moments.  As the quantity $D_q$ is the most accessible to analytical methods, and the most widely studied in the literature, we concentrate on it.   We have nevertheless checked that our results are similar for both quantities.

\section{Multifractal exponents for WP and eigenvectors} \label{results}
\begin{figure}[h!]
\includegraphics[width=0.95\linewidth,angle=0]{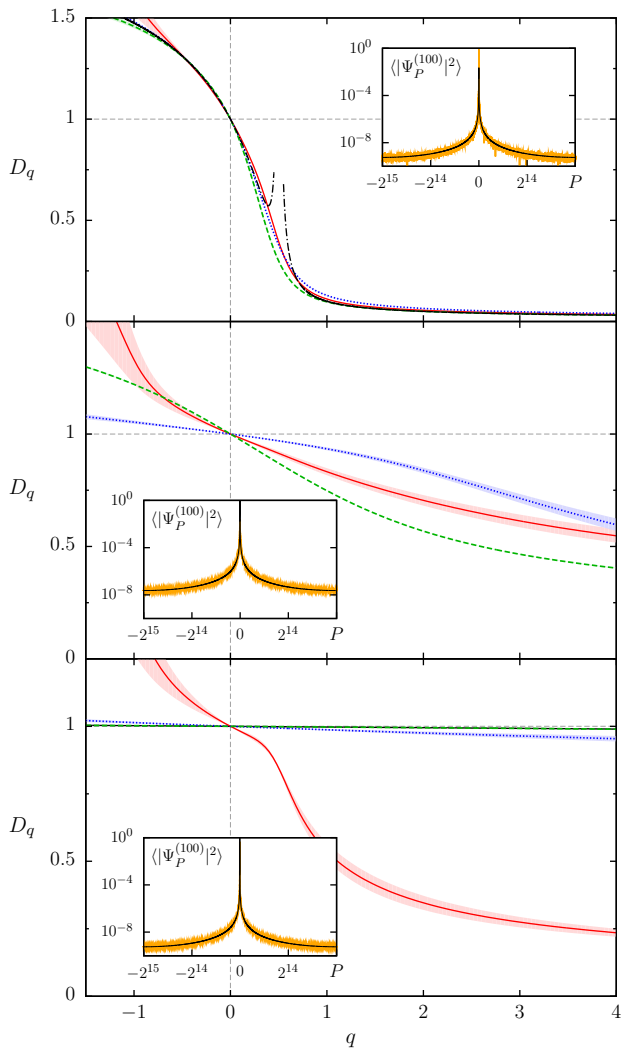}
\caption{(Color online) Multifractal exponents $D_q$ vs $q$ for
$\gamma=0.05$ (top), $\gamma=0.5$ (middle) and $\gamma=0.95$ (bottom). Red
solid curves: WP at $t=100$; blue dotted curves:
$t=\infty$ limit; green dashed curves: eigenvectors of (\ref{ruij}); black dotted-dashed curves: analytical theory for eigenvectors (see text).  Shaded areas indicate standard error in the least-square fit. Insets: average WP (initially localized at $P=0$)(dotted yellow curve) 
and analytical formula 
(\ref{bigres}) (black solid curve). The $D_q$ of
WP have been extracted from $1000$ random phase realizations of size
$N=2^{16}$ by the BC method applied on the $2^{15}$ central
components with box sizes from $1$ to $64$. The $D_q$ in
the $t=\infty$ limit and for eigenvectors have been extracted from $2^{13}$
vectors of size $2^{12}$ by the BC method with box sizes from
$8$ to $512$. All quantities on the figure are dimensionless.
\label{fig1}}
\end{figure}

We now turn to the discussion of multifractal exponents $D_q$ for WP
in different regimes of $\gamma$.  Studies of eigenvectors of the map (\ref{ruij}) have shown \cite{BogGir11,BogGir12} that their multifractality is the strongest close to $\gamma=0$ and the smallest for $\gamma$ close to nonzero integers.
In Fig.~\ref{fig1} we show the exponents $D_q$ for WP at $t=100$ and 
 $t=\infty$  and for eigenvectors in different regimes.  As long as the WP remains localized, the multifractal exponents are extracted from scales smaller than the
typical WP size. Indeed, above this scale, the weight is concentrated only in one box
and does not depend anymore on the box size (see insets of Fig.~\ref{fig0} showing the saturation of the moments). For very small times there are very few scales from which to extract the exponents, which may affect the precision.

The comparison of the different curves in Fig.~\ref{fig1} (top) shows that the regime of small $\gamma$ corresponds to a strong multifractality
 of both WP and eigenvectors.  
In this regime, it is possible to use a specific
perturbative approach in order to obtain multifractal properties
of eigenvectors \cite{BogGir12}. It predicts that for $q>1/2$ the multifractal
exponent is $D_q=2\gamma \frac{\Gamma(q-1/2)}{\sqrt{\pi}\Gamma(q)}$, while for
$q<1/2$ it is $D_q=\frac{2q-1}{q-1}+2\gamma \frac{\Gamma(1/2-q)}{\sqrt{\pi}(q-1)\Gamma(-q)} $. The results displayed in Fig.~\ref{fig1} (top) show that this formula is also quite close to the multifractal exponents of WP, even for
values of $\gamma$ as high as $0.05$.  Our explanation is that the
eigenvectors and initial WP being both very localized, only a few eigenvectors contribute to the WP. Thus in this regime the multifractal exponents of the WP yield a direct information on those of eigenvectors. We also note that the $t=\infty$ limit is reached already for $t=100$.  

When $\gamma$ goes further away from zero, 
the multifractality of eigenvectors decreases. As can be seen from the numerical data displayed
 in  Fig.~\ref{fig1} (middle) for $\gamma=0.5$, the multifractality of WP also becomes weaker.  In this regime the multifractal exponents for $t=100$ and
$t=\infty$ are quite close, showing that the $t=\infty$ limit is reached quite fast. This is  all the more remarkable since as shown in the insets of Fig.~\ref{fig1} top and middle, the WP at $t=100$ remains on average quite localized, while the envelope at $t=\infty$ is flat (data not shown).  In this regime, the asymptotic limit is thus quickly reached, and corresponds to a multifractality weaker than for eigenvectors.
Our interpretation is that as eigenvectors are more delocalized than for $\gamma\approx 0$, 
the initial WP has significant components on more
eigenvectors, which lead to an overall decrease of the multifractality as time evolution mixes these eigenfunctions.

When $\gamma$ increases and gets close to nonzero integer $k$, eigenvectors
display weak  multifractality.  This  can be derived analytically since 
the perturbative approach yields the expression $D_q=1-(\gamma -k)^2 q/k^2 $ \cite{BogGir12}.
In this regime the WP at $t=\infty$ are also weakly multifractal, as can be seen from the numerical data displayed
 in  Fig.~\ref{fig1} (bottom). The two curves are quite close but in this regime of very weak multifractality eigenvectors are slightly less multifractal than the $t=\infty$ limit.
For $t=100$ the multifractality of WP is quite different from the asymptotic one at $t=\infty$.  We will come back to this point below.

\begin{figure}[h!]
\includegraphics[width=0.95\linewidth,angle=0]{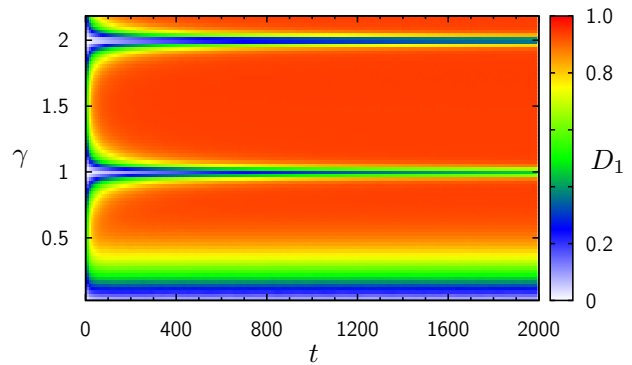}
\caption{(color online). Density plot of $D_1$ as a
function of $\gamma$ and time $t$. Colors denote multifractality strength from
white (strong) to red/gray (weak). The exponents $D_1$ have been extracted from
$100$ random phase realizations of size $N=2^{13}$ by the BC method
with box sizes from $1$ to $64$. All quantities on the figure are dimensionless, apart from the time t, in unit of period $T$.
\label{fig2}}
\end{figure}

\begin{figure}[h!]
\includegraphics[width=0.95\linewidth,angle=0]{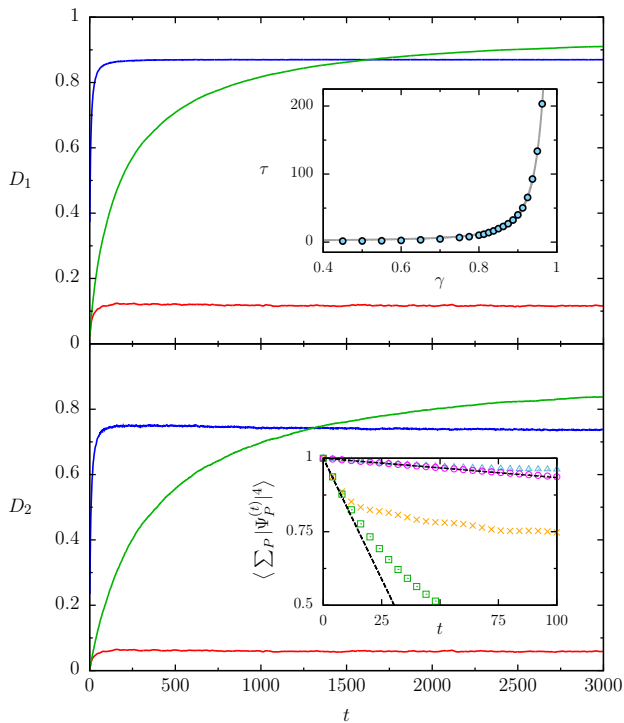}
\caption{(color online) Multifractal exponents $D_1$ (top) and $D_2$ (bottom) vs $t$, extracted from $1000$ random phase
realizations of size $N=2^{13}$ by the BC method with box sizes
 from $1$ to $64$. Red/bottom curves: $\gamma =0.05$; blue/top curves: $\gamma =0.5$; green/middle curves: $\gamma =0.95$. Upper inset : time $\tau$ (defined by $D_1(\tau)=D_{1,\mathrm{as}}/2$) vs $\gamma$. Here $D_{1,\mathrm{as}}$
is the mean value of $D_1(t)$ in the time interval $[3000,5000]$.
The solid curve is the best fit of the form
$\exp\left(A/(B-\gamma)\right)$ with $A= 0.749$ and $B = 1.103$. Lower inset:  mean second moment vs time for $\gamma=0.95$
(green squares), $0.05$ (orange crosses), $0.99$ (magenta circles) and $0.01$
(blue triangles).  Black dashed lines is the analytical formula for the second moment (see text). All quantities on the figure are dimensionless, apart from the time t, in unit of period $T$.
\label{fig3}}
\end{figure}

The global picture for the multifractality of WP is summarized 
in Fig.~\ref{fig2}, which displays $D_1$ as a function of time and $\gamma$.
The three regimes can be clearly distinguished, both in the average multifractality and in the speed with which the asymptotic regime is reached.  In order
to shed more light on the way multifractality evolves with time we show in
Fig.~\ref{fig3} the time evolution of $D_1$ and $D_2$ for three different values of $\gamma$.  While the asymptotic regime is reached very quickly for
$\gamma\approx 0$, the rate of convergence decreases with $\gamma$, as can be
checked more quantitatively with the data shown in the top inset.  Our interpretation of this phenomenon is the following; as $\gamma$ increases, we have already noted that the initial WP has significant components on more and more eigenvectors. One can relate the time at which the asymptotic regime is reached to
the inverse of the average spacing between eigenphases corresponding to these eigenfunctions which contribute.  For $\gamma\approx 0$ the eigenphases are close to the random variables $\Phi_P$ and  only few eigenvectors contribute, leading to a very large mean spacing and thus a very short convergence time.  As $\gamma$ increases, this mean spacing decreases and the convergence time increases. 

A perturbative method similar to the one used to obtain (\ref{bigres})
can be developed for the second moment of WP for $\gamma$ close to integers. It leads to  $\langle \Sigma_P |\Psi_P^{(t)}|^4 \rangle^{-1} \approx 1+2\pi^2 \epsilon^2 (N^2-1)/(3N^2) t$, valid for $t \ll 3/(2\pi^2 \epsilon ^2)$; data in Fig.~\ref{fig3} (lower inset) confirm the increasing range of validity of this formula when $\gamma$ gets closer to integers. It predicts $D_2 \approx 0$, compatible with the asymptotic limit for $\gamma \approx 0$, but not for $\gamma$ close to other integers.  This analysis therefore further confirms that as $\gamma$ gets closer to nonzero integers (weak multifractality regime), the asymptotic behavior should take longer and longer times to appear.

\section{conclusion}
We have studied the multifractality of individual WP in a periodically kicked system through
a combination of numerical and analytical works. We have compared different methods to  define and measure it, and assessed their usefulness, singling out the BC method as the most efficient in this context. The multifractality of WP was shown to typically decrease with time until it reaches an asymptotic limit, which corresponds to the model with randomized eigenvalues. This asymptotic multifractality is different from the one of eigenvectors more commonly studied, but is related to it. The rate at which the asymptotic limit is reached can also be related to the multifractality of eigenvectors. Although the model we used stems from mathematical studies, we think our results should be applicable to other models and in particular can guide the analysis of experimental situations.  

\acknowledgments We thank CalMiP for access to its supercomputers,
the FNRS, the NEXT project ENCOQUAM and the University Paul Sabatier for funding (OMASYC project). I.G.M. received support from ANCyPT grant PICT 2010-1556 
and from CONICET grant PIP 114-20110100048.


\begin{thebibliography}{99}
\bibitem{turbulence} C.~Meneveau and K.~R.~Sreenivasan,  Phys. Rev. Lett. {\bf 59}, 1424
(1987); J.-F.~Muzy, E.~Bacry and A.~Arneodo, Phys. Rev. Lett. {\bf 67}, 3515 (1991).
\bibitem{stock} B.~B.~Mandelbrot, A.~J.~Fisher and L.~E.~Calvet,  Cowles Foundation Discussion Paper No. 1164 (1997).
\bibitem{cloud} S.~Lovejoy and D.~Schertzer,
Journal of Geophysical Research {\bf 95}, 2021 (1990).
\bibitem{trans} C.~Castellani and L.~Peliti, J. Phys. A {\bf 19}, L429 (1986); J.~Bauer, T.~M.~Chang and J.~L.~Skinner, Phys. Rev. B {\bf 42}, 8121 (1990); M.~Schreiber and H. Grussbach, Phys. Rev. Lett. {\bf 67} 607 (1991); W.~Pook and M.~Janssen, Z. Phys. B {\bf 82}, 295 (1991); B.~Huckestein, B.~Kramer and L.~Schweitzer, Surface Science {\bf 263}, 125 (1992).
\bibitem{mirlin2000}
A.~D.~Mirlin, Phys. Rep. {\bf 326}, 259 (2000).
\bibitem{mirlinRMP08} F.~Evers and A. D. Mirlin, Rev. Mod. Phys. 
{\bf 80}, 1355 (2008).
\bibitem{romer} A.~Rodriguez, L.~J.~Vasquez and R.~A.~R\"omer, Phys. Rev. Lett. {\bf 102}, 106406 (2009).
\bibitem{huckenstein} B.~Huckestein, Rev. Mod. Phys. {\bf 67}, 357 (1995).
\bibitem{PRBM} A.~D.~Mirlin, Y.~V.~Fyodorov, F.-M.~Dittes, J.~Quezada, and
  T.~H.~Seligman, Phys. Rev. E {\bf 54}, 3221 (1996); J.~A.~Mendez-Bermudez, A.~Alcazar-Lopez and I.~Varga, Europhysics Letters {\bf 98} 37006 (2012).
\bibitem{ossipov} Y.~V.~Fyodorov, A.~Ossipov and A.~Rodriguez, J. Stat. Mech. L12001 (2009).
\bibitem{indians1} N.~Meenakshisundaram and A.~Lakshminarayan,  Phys. Rev. E {\bf 71}, 065303(R) (2005).
\bibitem{indians} J.~N.~Bandyopadhyay, J.~Wang and J.~Gong, Phys. Rev. E {\bf 81}, 066212 (2010).
\bibitem{garciagarcia} A.~M.~Garc\'{\i}a-Garc\'{\i}a and J.~Wang,
Phys. Rev. Lett. {\bf 94}, 244102 (2005).
\bibitem{interm} E.~B.~Bogomolny, U.~Gerland and C.~Schmit,  
Phys. Rev. E {\bf 59}, R1315 (1999); E.~B.~Bogomolny, O.~Giraud and C.~Schmit, Phys. Rev. E
{\bf 65}, 056214 (2002).
\bibitem{giraud} O.~Giraud, J.~Marklof and S.~O'Keefe, J. Phys. A
{\bf 37}, L303 (2004).
\bibitem{bogomolny} E.~B.~Bogomolny and C.~Schmit,
Phys. Rev. Lett. {\bf 93}, 254102 (2004).
\bibitem{MGG} J.~Martin, O.~Giraud and B.~Georgeot,  Phys. Rev. E {\bf 77}, 035201(R) (2008).
\bibitem{MGGG} J.~Martin, I.~Garc\'{\i}a-Mata, O.~Giraud and B.~Georgeot,  Phys. Rev. E {\bf 82}, 046206 (2010).
\bibitem{BogGirPRL}E.~Bogomolny and O.~Giraud, Phys. Rev. Lett. {\bf 106}, 044101 (2011).
\bibitem{BogGir11}E.~Bogomolny and O.~Giraud, Phys. Rev. E {\bf 84}, 036212 (2011). 
\bibitem{BogGir12}E.~Bogomolny and O.~Giraud, Phys. Rev. E {\bf 85}, 046208 (2012). 
\bibitem{billes} S.~Faez, A.~Strybulevych, J.~H.~Page, A.~Lagendijk and B.~A.~ van Tiggelen,  Phys. Rev. Lett. {\bf 103} 155703 (2009).
\bibitem{richard}   A.~Richardella, P.~Roushan, S.~Mack, B.~Zhou, D.~A.~Huse, D.~D.~Awschalom and A.~Yazdani, Science {\bf 327}, 665 (2010).
\bibitem{cold}G.~Lemari\'e, H.~Lignier, D.~Delande, P.~Szriftgiser and J.-C.~Garreau, Phys. Rev. Lett. {\bf 105}, 090601 (2010); Y.~Sagi, M.~Brook, I.~Almog and N.~Davidson, Phys. Rev. Lett. {\bf 108}, 093002 (2012).
\bibitem{chalker} J.~T.~Chalker and G.~J.~Daniell,  Phys. Rev. Lett. {\bf 61}, 593 (1988).
\bibitem{schweitzer} B.~Huckestein  and L.~Schweitzer,  Phys. Rev. Lett. {\bf 72}, 713 (1994).
\bibitem{guarneri} I.~Guarneri and G.~Mantica,  Phys. Rev. Lett. {\bf 73}, 3379 (1994).
\bibitem{geisel} R.~Ketzmerick, K.~Kruse, S.~Kraut and T.~Geisel,  Phys. Rev. Lett. {\bf 79}, 1959 (1997).
\bibitem{raizen} J.~Zhong, R.~B.~Diener, D.~A.~Steck, W.~H.~Oskay, M.~G.~Raizen, E.~W.~Plummer, Z.~Zhang and Q.~Niu, Phys. Rev. Lett. {\bf 86}, 2485 (2001).
\bibitem{schreiber} H.~Grussbach and M.~Schreiber, Phys. Rev. B {\bf 48}, 6650 (1993).
\bibitem{grecs} S.~N.~Evangelou and D.~E.~Katsanos, J.~Phys.~A {\bf 26}, L1243 (1993).
\bibitem{ruijsc} S.~N.~M.~Ruijsenaars and H.~Schneider, Ann. Phys. {\bf 170}, 370 (1986).
\bibitem{arneodo} A.~Arneodo in {\em Wavelets: Theory and Applications}, G.~Erlebacher, M.~Y.~Hussaini and L.~M.~Jameson, Eds (Oxford University Press, 
         New York, 1996).
\bibitem{kantel} J.~W.~Kantelhardt, H.~E.~Roman and M.~Greiner, Physica A {\bf 220}, 219 (1995).
\bibitem{nous}  I.~Garc\'{\i}a-Mata, O.~Giraud and B.~Georgeot, Phys. Rev. A {\bf 79}, 052321 (2009).
\end{thebibliography}
\end{document}